\documentclass[10pt]{article}
\usepackage{epsfig}

\begin{document}

\begin{center}
{\Large \bf Search for new heavy resonances at the LHC} \\
\vspace{4mm}

I.~Golutvin, \underline{E.~Rogalev}, M.~Savina, S.~Shmatov \\
Jonit Institute for Nuclear Research, Dubna, Russia \\
\end{center}

\begin{abstract}
In this article we carry out an analysis of LHC potential to
search for new dimuon resonance states from extended gauge models
and the Randall-Sundrum scenario of TeV-scale gravity.
\end{abstract}

\section{Introduction}

    The Standard Model (SM) had been tested in many experiments at LEP, SLC
and Tevatron with a high accuracy. In particular, the yield of
lepton pairs produced mainly via Drell-Yan processes, i.e.
quark-antiquark annihilation by exchange of photons or Z bosons,
is predicted by the SM with a per mille precision. So far, the
experimental data have shown no significant deviations from SM
predictions for the Drell-Yan continuum up to TeV-energy-scale.

The high-order calculations of lepton pair production cross
section in the mass region of $ 0.1 \div 0.8 $
TeV/$ c^{2} $ are in good agreement with LEP and D0 data \cite{D0,OPAL}.
At present, however, there are many theoretical attempts to extend
the bounds of the SM in order to reach unification of strong and
electroweak interactions or remove an arbitrariness in values of
coupling constants in some other way and also to pull through
known disadvantages of SM like the mass hierarchy problem,
CP-violation problem etc. Supersymmetry is the most popular
theoretical extension of the SM, however, some other alternatives
also exist. Between them one can consider extended gauge models
(what contain extended sector of gauge bosons in comparison with
the SM). Consideration of symmetry groups wider than in the SM and
appearance of extra gauge bosons is common feature of various
left-right symmetric models, all variants of Grand Unification
theories and models of composite gauge bosons
\cite{Cvetic95,Hewett89}. For all of these cases new vector
bosons, neutral $Z^{\prime}$ and charged $W^{\prime}$, would
appear at the mass scale of order of one TeV/$c^{2}$ what can be
observed at the LHC. Another alternative way to go beyond the SM
can be to consider so-called TeV-scale gravity models, as given by
brane world scenarios with large or compact extra spatial
dimensions (LED) \cite{ADD,Randall99}. These models give in
particular a set of new particle - massive Kaluza-Klein modes of
graviton what can interact with the usual matter and contribute to
the SM processes causing some new interesting physics at the TeV
energy scale (for more details and a phenomenology review see, for example
\cite{Rubakov01,Kazakov04}).

    In our consideration we touch on only one of the possible LED scenarios, the
Randall-Sundrum (RS1) approach \cite{Randall99}. The standard setup for this approach
contains two three-branes embedded into the external curve five-dimensional
space, one brane is with negative tension (our world with all
usual SM fields) and another one is with positive tension
(Planckian brane) what is needed to reproduce the SM mass
hierarchy. The usual assumption is that all the SM fields and
fields from the planckian sector are confined on corresponding
branes and effectively four-dimensional. Graviton is only really
multidimensional field what can travel freely through the whole
space (the bulk space). Because of nonzero curvature of the bulk
space (what is represented by a slice of the five-dimensional AdS
space bounded by two branes) and the specific space-time geometry
new very interesting physics can be realized at our brane. In
particular, in any model with compact extra spatial dimensions
each matter field will have an infinite tower of massive
excitations called Kaluza-Klein modes (four-dimensional
"projections" of multidimensional fields from our point of view).
These modes can be visible at energy above the fundamental theory
scale, say, from energies above one or a few TeV. Then our usual
massless four-dimensional graviton can be treated as the zero mode
from the KK-tower. But the distinctive feature of particular RS1
phenomenology is that excited massive graviton states are strongly
coupled to ordinary particles (not suppressed below the planckian
scale like for ordinary graviton in usual description of gravity)
and can contribute significantly to the SM processes above the
fundamental scale. Mass splitting between KK-modes of graviton is
of order ${\Delta}m{\sim}ke^{-kr_{c}}$ \cite{Rubakov01}, and
values of masses start from the TeV scale. So, at least the first
Kaluza-Klein mode (called RS1 graviton below) can in principle be
observed at the LHC as individual heavy (quite narrow or not)
resonance.

At this point it is interesting to note that there exist also some realizations of
TeV-scale gravity models where fields other than graviton can have Kaluza-Klein modes,
see for example a number of papers \cite{Antoniadis90}
{\footnote{Authors are grateful to I.Antoniadis who has drawn our
attention to this issue}}.
In the papers interesting scenarios based on superstring theory were proposed in which
all gauge bosons and higgses as far as their superpartners had KK-excitations.
We will not touch on such scenarios in our further
discussion but we would stress that study of phenomenological consequences of the existence
of KK-modes of $Z^0$ or photon can be done exactly in the same spirit as
it is presented in subsections 2.1 -- 2.2 below for $Z^{\prime}$ including an asymmetry analysis.

For both of the conceptions
considered, $Z^{\prime}$ and RS1 graviton, a width of predicted
resonances is not fixed but it can vary widely depending on the
model parameters. It implies that these states can appear as
individual resonances or can affect the high-$p_{T}$ lepton pair
continuum leading to an excess of Drell-Yan production. Thus, the
distinctive experimental signature for these processes is a pair
of well-isolated high-$p_{T}$ leptons with opposite charges coming
from the same vertex.

    Such the measurements can be performed at the both of LHC experiments, ATLAS
and CMS, which are expected to be able to trigger and identify
hard muons with a transverse momentum up to several TeV. The
ability of the LHC experiments to detect RS1 graviton in the
dielectron mode was investigated as described in
Ref.~\cite{RS_CMS}. In this paper, we present the analysis of the
LHC discovery limit on $Z^{\prime}$ and RS1 gravitons in dimuon
mode in assumption of the CMS acceptance.

\section{Extra gauge bosons}
\subsection{Signal and background simulations}

    The signal simulation is done for the parton subprocess $q\bar{q}{\rightarrow}Z^{\prime}$ in the
leading order of QCD without higher order corrections. To generate
$Z^{\prime}$ boson and its decay to a muon pair as well as the
relevant background events, the PYTHIA 6.217 package \cite{PYTHIA}
with the CTEQ5L parton distribution function was used. There exist
a number of possible non Standard Model scenarios which predict
appearance of heavy neutral $Z^{\prime}$ and/or charged
$W^{\prime}$ gauge bosons (for review see \cite{Cvetic95}). But
only following $Z^{\prime}$ models were used in our analysis:

\begin{enumerate}
  \item The Left-Right model (LR) \cite{Huitu97} based on the electroweak gauge group symmetry
$SU(2)_{L} \times SU(2)_{R} \times U(1)_{B-L}$ with an additional
left-right symmetry (here, $B$ and $L$ are the baryon and lepton
numbers) with default PYTHIA couplings which are the same for both
left- and right-handed type of fermions and are set to the same
values as in the SM, $g_{L}$ = $g_{R}$ = 0.64. The number of extra
fermion generations is equal to three.

  \item $Z^{\prime}_{\chi}-$, $Z^{\prime}_{\eta}-$, and $Z^{\prime}_{\psi}$ -
models which naturally arise as a result of sequential
breaking of SO(10) or $E_6$ symmetry group in Grand Unified
Theories (GUT) \cite{Hewett89}: $E_6 \rightarrow SO(10) \times
U(1)_{\psi} \rightarrow SU(5)\times U(1)_{\chi} \times
U(1)_{\psi}$ $\rightarrow SM{\times}U(1)_{{\Theta}_6}$. The linear
combination of the hypercharges under the two groups
$U(1)_{\chi}{\times}U(1)_{\psi}$ gives the charge of the lightest
$Z^{\prime}$ at the symmetry-breaking energies $Z^{\prime}$ =
$Z^{\prime}_{\chi}cos({\Theta}_{E_6}) +
Z^{\prime}_{\psi}sin({\Theta}_{E_6})$. Numerical values of the
couplings for these models are taken from Ref.~\cite{Rosner87}.

  \item For Monte Carlo studies the sequential standard model (SSM)
\cite{Altarelli89} was also used in which heavy bosons
($Z^{\prime}$ and $W^{\prime}$) are assumed to couple only to one
fermion type (left) with the same parameters (couplings and total
widths) as for ordinary $Z^{0}$ and $W^{\pm}$ in the Standard
Model.
\end{enumerate}

\begin{figure}[thb]
\vspace{-2.5cm}
\centerline{\psfig{figure=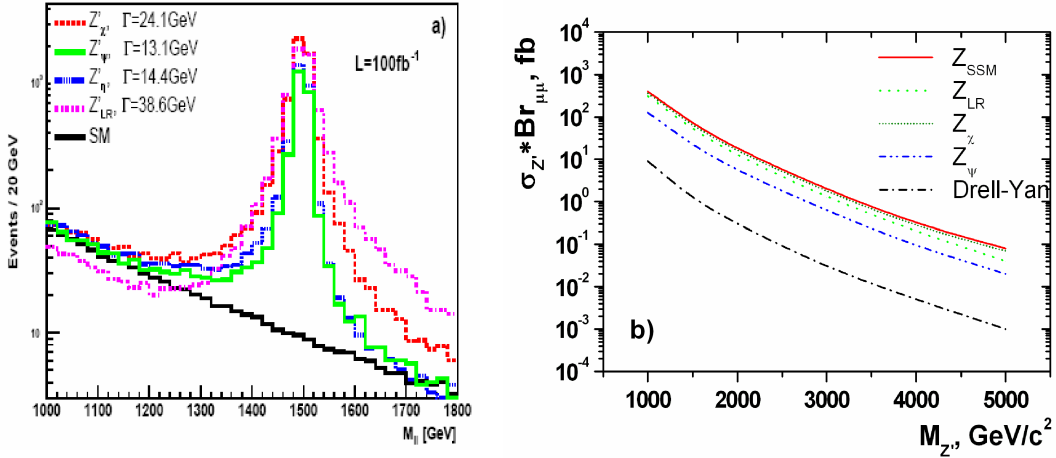,width=18cm}} \vspace{-18.cm}
\caption{Invariant mass distribution of dimuon pairs for extended
gauge models and SM (a) (from \cite{Dittmar04}). Cross section of
dimuon produced from $Z^{\prime}$ decay in dependence on a mass
scale (b). Also the SM prediction for DY is presented (the lower
curve)} \label{fig1}
\end{figure}

    The non-reducible background considered is the SM Drell-Yan process
$pp{\rightarrow}Z/{\gamma}{\rightarrow}{\mu}^{+}{\mu}^{-}$ which
gives nearly 95 $\%$ of the Standard Model muon continuum.
Contributions from other reaction (vector boson pair production
$ZZ$, $WZ$, $WW$, quark-antiquark production $t\bar{t}$ etc) are
very small and neglected in this study. In the SM the expected
number of dimuon events is not very large and the $Z^{\prime}$
resonance peak exceeds the background by above a factor ten
(Fig.~\ref{fig1}, the left plot). The mass dependence of $Z^{\prime}$
production cross secrion times on dimuon branching are given in
Fig.~\ref{fig1} (the right plot). For comparison the cross section for
standard Drell-Yan muon pair production as a function of their
invarinat masses is also presented on the same figure.

\subsection{$Z^{\prime}$ discovery limits}

    Event samples for seven mass values were generated with above-mentioned
model parameters. To take into account the detector response the
parametrization of a muon momentum ($p$) resolution,
4$\%$/$\sqrt{p/TeV}$ \cite{Muon TDR}, was used. The dimuon is
accepted when both decay muons are within detector system covering
the pseudorapidity region of $|$${\eta}$$|$ ${\leq}$ 2.4. In
addition, the cut $p_{T}{\geq}20$ GeV/c was applied on each muon.
No cuts were made on isolation of muons in the tracker and the
calorimeter. The total efficiency dimuon selection,
${\varepsilon}$, is about $83{\div}91$ $\%$.

\begin{figure}[hbt]
\vspace{-1.5cm} \centerline{\psfig{figure=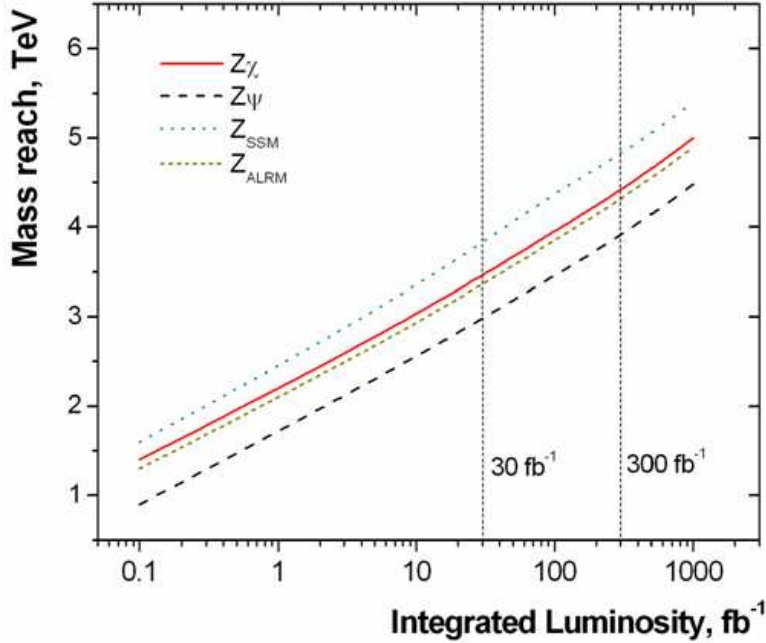,width=13cm}}
\vspace{-8.5cm} \caption{Upper limit for $Z^{\prime}$ mass with
statistical significance of the signal of 5${\sigma}$}
\label{fig2}
\end{figure}

    To estimate the $Z^{\prime}$ discovery limit expected significance of the signal,
 $S_{c12}$=$\sqrt{S+B}$-$\sqrt{B}$, was computed, where $S$ is the number of signal
events passed through all kinematics cuts and $B$ is the number of background
events. The discovery limits for a five ${\sigma}$ signal are
presented in Fig.~\ref{fig2}.

Detection of the $Z^{\prime}$ peak itself and precise measurement of its mass
and width does not allow an underlying theoretical model
describing the $Z^{\prime}$ to be identified. To test a helicity
structure of the boson and discriminate between a number of
$Z^{\prime}$ models, a leptonic forward-backward asymmetry can be
used (see, for example, \cite{Langacker84}). The asymmetry is
defined as the ratio $A_{FB} = \frac{(F-B)}{(F+B)}$ , where $F$
and $B$ are the numbers of events in the forward and backward
directions, respectively. Forward (backward) is defined as the
hemisphere with $cos({\Theta}) > 0$ ($cos({\Theta}) < 0$), where
${\Theta}$ is an angle between the outgoing negative lepton and
the quark $q$ in the $q\bar{q}$ rest frame. Such a definition
assumes that the original quark direction is known, but this is
not the case for the pp-experiment. In Ref.~\cite{Dittmar97},
however, it was shown that it is possible to identify the quark
direction with the boost direction of the dimuon system with
respect to the beam axis.

\begin{figure}[hbt]
\vspace{-2.0cm}
\centerline{\psfig{figure=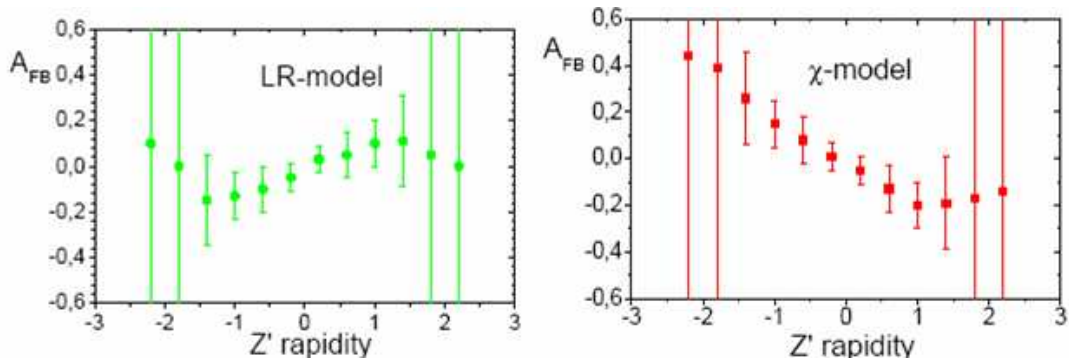,width=18cm}} \vspace{-19.3cm}
\caption{Forward-backward asymmetry for different $Z^{\prime}$ models.} \label{fig3}
\end{figure}

    One of the features of the asymmetry, $A_{FB}$, is the distinctive rapidity-dependence
for different $Z^{\prime}$ models. Such a dependence is shown in
Fig.~\ref{fig3} for the $Z^{\prime}_{LR}$ (the left picture) and the
$Z^{\prime}_{\chi}$ (the right picture) under assumption of a mass
$M_{Z^{\prime}}$ = 2.0 TeV/$c^{2}$, for 100 $fb^{-1}$ of an
integrated luminosity.

Logic of such an analysis looks quite consequent but it is not complete in general
as another one possibility is not accounted by it.
As it was yet pointed out (see comments on the theme in Introduction above)
one can image a situation when we will see a resonance at collider  with a mass about TeV and quantum numbers
of gauge boson what can be in principle not $Z^{\prime}$ from the extended gauge sector but Kaluza-Klein
excitation $Z_{KK}$ coming from models with extra spatial dimensions. The only way to distinguish these quite
different cases when will be to look for other KK-excitations of other gauge bosons (photon, gluon) what must
exist in the presence of extra dimensions \cite{Antoniadis90}.

\section{RS1 graviton}

\subsection{Signal and background simulations}

\begin{figure}[htbp]
\vspace{-0.5cm}
\centerline{\psfig{figure=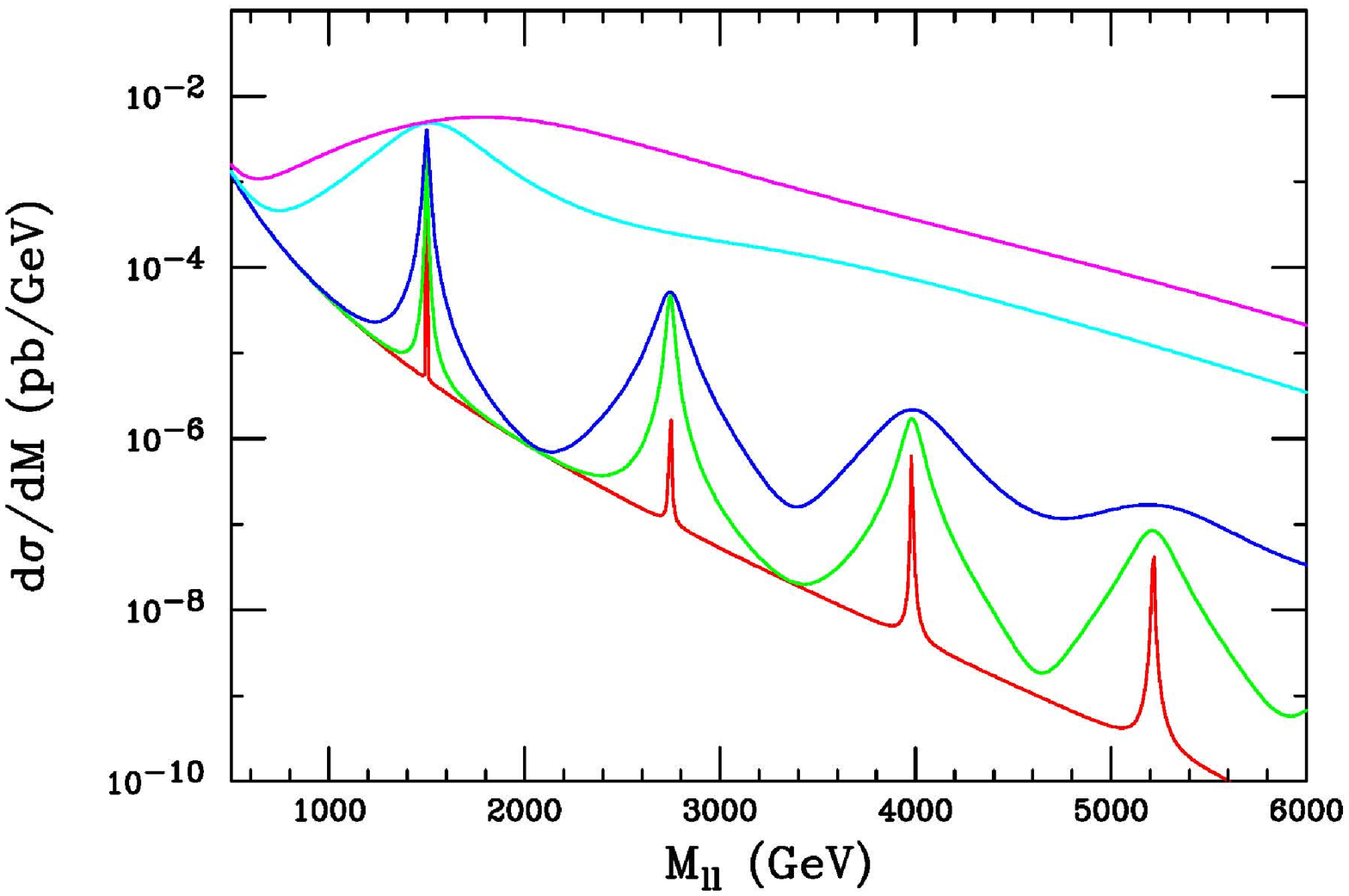,height=11.5cm,angle=90}}
\vspace{-1.0cm} \caption{Invariant mass distributions for
Drell-Yan process for the LHC. From top to bottom, the curves are
for coupling values $c$ = 1, 0.5, 0.1, 0.05, 0.01 (taken from
\cite{Davoudiasl01})} \label{fig4}
\end{figure}

As it was mentioned above the RS1 model predicts massive KK-modes
of graviton ($m_{KK} \approx $ a few TeV) which can been observed
in experiments as heavy narrow resonance states (Fig.~\ref{fig4}). It is
necessary to note here that an effect from KK-gravitons can also
been extracted from changing of Higgs boson rates because of
mixing of them with radion (see Ref.~\cite{Wells02})

    The ability to test experimentally RS1 scenario predictions depends on the
model parameter $c = k/M_{Pl}$ which controls coupling of graviton
to the ordinary particles and a width of the resonance
${\Gamma}$${\sim}$${\rho}m_{0}c^{2}$; where the constant ${\rho}$
is determined by the number of open decay channels. Results of the
combined analysis in the RS1 scenario \cite{Davoudiasl01} show
that a value of the dimensionless coupling constant $c$ and a
corresponding value of graviton mass is restricted due to the
experimental Tevatron data and theoretical constraints to assure
the model hierarchy (${\Lambda}_{\pi}$ $<$ 10 TeV). These
limitations lead, in particular, to the conclusion that the
constant $c$ can not be less than 0.027, for the graviton mass of
one TeV/$c^{2}$, and less than 0.1, for the mass of 3.7
TeV/$c^{2}$. We operated in our consideration the range for
coupling constant $0.01 {\leq} c {\leq} 0.1$. Graviton resonances
can be produced via quark-antiquark annihilation $q\bar{q}
{\rightarrow} G_{KK}$ as well as gluon-gluon fusion $gg
{\rightarrow} G_{KK}$. The first of these processes is identical
to the Standard Model s-channel exchange by an intermediate
$\gamma^{\star}$ or $Z$ vector boson, while the second one has no
SM analogue. Other partonic sub-processes are also possible,
$gg{\rightarrow} gG_{KK}$, $q\bar{q} {\rightarrow} gG_{KK}$,
 $gq {\rightarrow} qG_{KK}$, which form real graviton production via the
$t$-channel exchange (graviton emission).
To simulate both real and virtual graviton production in the
proton-proton collisions at 14 TeV center-of-mass energy, PYTHIA
6.217 was used in which the RS1 scenario was implemented with
CTEQ5L parton distribution functions. The graviton production
cross section for all five possible diagrams is presented in Table~\ref{RS1_crs}.
Here, two extreme possibilities for model parameter $c$ were
considered. The first one was the most optimistic scenario when $c$
is equal to 0.1. The second number in each row (in brackets)
corresponds to the most difficult for experimental observation
case when the coupling value $c$ = 0.01. The majority (at least
50${\div}$60 $\%$ depending on a mass) of gravitons was produced
in the process of gluon-gluon fusion with real graviton emission,
whereas virtual graviton production added only up to 15 $\%$ of
the total cross-section. The Standard Model background for this
channel was the same as for the $Z^{\prime}$ case.

\begin{table}[h,pt]
\caption{\label{RS1_crs} Leading-order cross section of
$\mbox{G}_{\mbox{\small KK}}$ production in fb. The CTEQ5L parton
distributions have been used.}
\vspace{-0.8cm}
\begin{center}
\item[]\begin{tabular}{@{}llll}
\hline
Mass, $\mbox{TeV}c^{-2}$                                              & $1.0$       &  $1.5$     & $3.0$ \\
\hline
$\mbox{q} \bar{\mbox{q}} \to \mbox{G}_{\mbox{\small KK}}$           & 129 (1.34)  & 23 (0.24)  & 0.633 (0.006)  \\
$\mbox{gg} \to \mbox{G}_{\mbox{\small KK}}$                         & 567 (5.33)  & 62 (0.53)  & 0.94 (0.004)  \\
$\mbox{q} \bar{\mbox{q}} \to \mbox{g} \mbox{G}_{\mbox{\small KK}}$  & 345 (3.29)  & 65 (0.64)  & 1.84 (0.017)  \\
$\mbox{qg} \to \mbox{q} \mbox{G}_{\mbox{\small KK}}$                & 599 (5.78)  & 72 (0.64)  & 1.05 (0.007)  \\
$\mbox{gg} \to \mbox{g} \mbox{G}_{\mbox{\small KK}}$                & 3350 (31.5) & 368 (3.32) & 4.98 (0.028)  \\
Total                                                               & 4990 (47.2) & 590 (5.38) & 9.45 (0.062) \\
\hline
\end{tabular}
\end{center}
\end{table}

\subsection{Detection of RS1-graviton resonance and discovery limits}

\begin{figure}[ht]
\vspace{1.9cm}
\begin{center}
\resizebox{11.5cm}{!} {\includegraphics{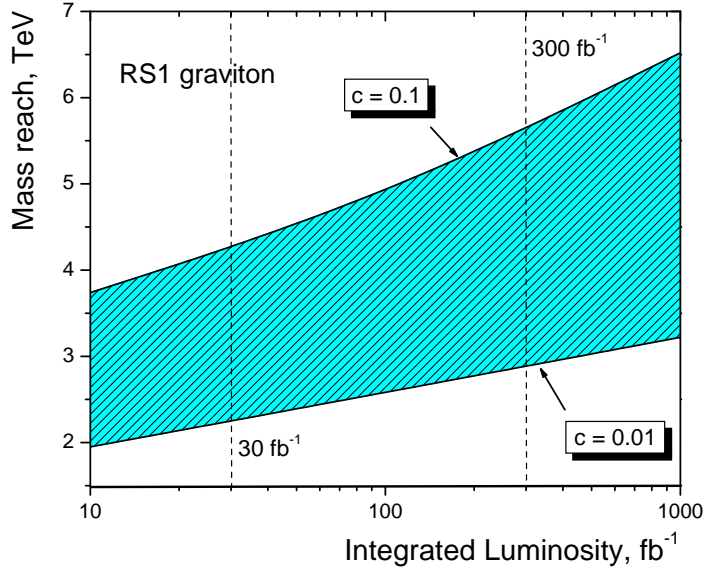}}
\vspace{-3.2cm}
\caption{Five ${\sigma}$ discovery limits for RS1
graviton decayed into dimuon pair.}
\label{fig5}
\end{center}
\end{figure}

    Simulation of the detector response for graviton decay into muon pair
is similar to the $Z^{\prime}$ case. To estimate the discovery
limit for RS1 graviton excitations the same procedure as for the
$Z^{\prime}$ case was applied. The cross section of $G_{KK}$
production and corresponding cross section limits to observe a
five ${\sigma}$ signal for various integrated luminosity cases are
presented in Fig.~\ref{fig5}. As shown in the figure, the LHC can test the
RS1 scenario in the whole range of the model parameter $c$ up to the
mass of 2.0 TeV/$c^{2}$ even at the low luminosity of 10
$fb^{-1}$. In the more favourable case with $c$ = 0.1 the
accessible mass region is extended up to 3.7 TeV/$c^{2}$. With 100
$fb^{-1}$ of integrated luminosity the reach increases up to 2.6
TeV/$c^{2}$ for $c$=0.01 and 4.8 TeV/$c^{2}$ for $c$=0.1.

     The direct comparison of results on an allowed region for $c$ with the data
of the Fig.~\ref{fig5} shows that the whole space of the RS1-model
parameters is accessible at the luminosity of 100 $fb^{-1}$, and
RS1 graviton can be discovered with the five ${\sigma}$
significance. These conclusions, however, are not definitive,
since initial theoretical constraints are very arbitrary.

\begin{figure}[hbt]
\vspace{-1.5cm} \centerline{\psfig{figure=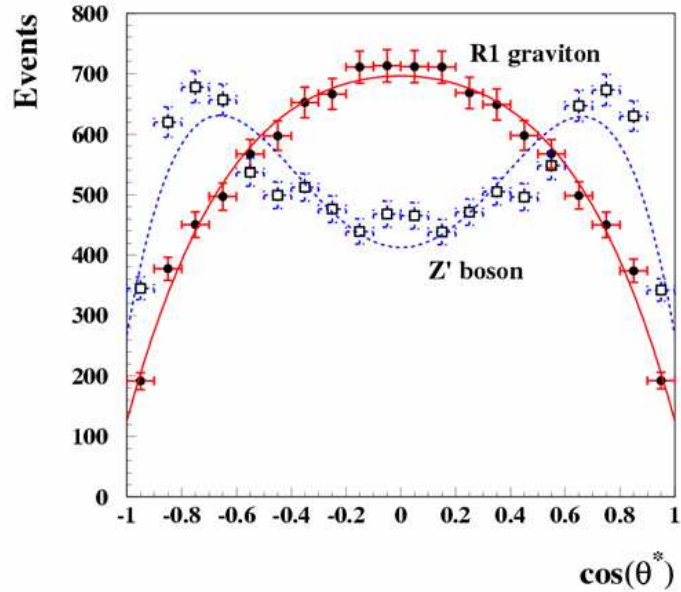,width=11.5cm}}
\vspace{-7.5cm} \caption{Angular distributions $cos({\Theta}*)$ of
muons from $G_{KK}$ (solid marker)and $Z^{\prime}$ (open box) decays}
\label{fig6}
\end{figure}

    Under the assumption that a new resonance state was observed at the LHC,
nature of this object should be understood in further analysis (in
principle, the resonance can come from the extended gauge sector
as well as from some version of extra dimension models). The major
difference between $Z^{\prime}$ and RS1 graviton should appear in
the $\mbox{cos}(\theta^{\star})$ distribution (where
$\theta^{\star}$ is the polar angle of muons in the center-of-mass
system of the dimuon pair) which is strongly spin dependent.
Course, these distributions will be distorted by acceptance cuts,
especially in the region of large angles, and expected theoretical
predictions will differ from experimental ones. Nevertheless Fig.~\ref{fig6}
shows the distinct difference between the spin-1 ($Z^{\prime}$)
and the spin-2 (RS1 graviton) curves obtained for muons after all
cuts. To ease the comparison, these plots were obtained for
resonance states with the same masses of 1.5 TeV/$c^{2}$ and
normalized to 6000 events that correspond to the approximate
integrated luminosity of 10 $fb^{-1}$ for graviton production with
$c$=0.1 and 100 $fb^{-1}$ for $Z^{\prime}$ boson.


\section{Summary}

In this work we present the discovery potential of $Z^{\prime}$
gauge bosons as well as RS1 gravitons in the muon channel at the
LHC experiments. The estimated discovery limit for $Z^{\prime}$ is
about $ 3.5 \div 4.0 $ TeV/$c^{2}$ depending on the couplings for
100 $fb^{-1}$ integrated luminosity. At the same luminosity the
first KK-mode of RS1 graviton can be observed up to the mass
values of 2.6 TeV/$ c^{2} $ and 4.8 TeV/$c^{2}$ for $c$ = 0.01 and
0.1, respectively. The angular distribution of muons in the final
state can be used to distinguish the spin-1 and the spin-2
resonance states, at least in the mass region up to 2.3
TeV/$c^{2}$. The different $Z^{\prime}$ models can be
distinguished (up to the mass value of 2.5 TeV/$c^{2}$) using the
leptonic forward-backward asymmetry. Further detailed studies are
required on the theoretical side in order to identify other
possible physics observables. This would be helpful to better
understand the LHC discovery conditions and limits.

Authors would like to thank E.~Boos, D.~Denegri, M.~Dubinin,
A.~Lanyov, and S.~Valuev for enlightening and helpful discussions.

E.R. is also grateful to Organizing Committee for hospitality and
very interesting scientific program of the Conference.

\end{document}